\documentstyle[prl,aps,preprint]{revtex}

\begin{document}

\title {A Possible Nanometer-scale Computing Device Based on an Adding
Cellular Automaton}
\author{Simon C. Benjamin$^{a)}$
and  Neil F. Johnson} 
\address{Physics Department, Clarendon
Laboratory, Oxford University, Oxford OX1 3PU, England} 

\maketitle

\begin{abstract}
We present a simple one-dimensional Cellular Automaton (CA) which has the property
that an initial state composed of two binary numbers evolves quickly into a final
state which is their sum. We call this CA the Adding Cellular Automaton (ACA). 
The ACA
requires only
$2N$ two-state cells in order to add any two $N-1$ bit binary numbers. The ACA could
be directly realized as a wireless nanometer-scale computing device - a possible
implementation using coupled quantum dots is outlined.

\end{abstract}  \vskip 0.2in  

\vskip 7cm
$^{a)}$ correspondence to: s.benjamin@physics.ox.ac.uk

\newpage
For more than thirty years the processing power associated with a single
silicon chip has increased exponentially. This growth results from the
continual increase in the number of transistors integrated into the chip.
State-of-the-art
chips now have features of the order of 1000 atoms wide. Before this number is
reduced by a single order, we may expect quantum effects to become sufficiently
strong that they must be explicitly allowed for in the design.  

One is naturally led to ask the
question, may there not be some wholly different computational architecture
that is more efficient (in some sense more `natural') in the quantum mechanical
regime? A number of theoretical studies (e.g.
\cite{domino,lines,Obermayer,parametron}) have proposed novel processing
systems of this kind.  These structures have a certain feature in common:
they consist of a large number of identical cellular units. Systems of this kind are
related to the mathematical idea of cellular automata.

The basic cellular automaton \cite{vonN} is a $D$-dimensional
lattice of identical cells, each of which is in one of a finite set of possible
internal states (called the `state alphabet'). The entire lattice, which may be
infinite or periodic, is updated in discrete time steps. On each update, each cell
assumes a new state determined by its own present state, and the state of a
certain set of local cells (the cell's `neighborhood'). For example, in a
one-dimensional CA a cell's neighborhood might consist of just the cell
itself and its immediate neighbors. The update rule, which specifies what internal
state to assume for each possible arrangement of states in the neighborhood, is
referred to as the `table'. The neighborhood, and the
`table', are the same for all the cells.

In the first part of this letter we introduce a one-dimensional CA 
designed to have the property that an initial state composed of two binary numbers
evolves into a final state which is their sum. We call this CA the
Adding Cellular Automaton (ACA). We have designed the ACA to include certain
features that make direct physical realization possible. We expect that there are
many physical systems capable of implementing the ACA scheme. In the second part of
the letter we provide one example, a modified version of the quantum dot
architecture proposed in Ref. \cite{Obermayer}.

The ACA has excellent efficiency; the addition of two
$N-1$ bit numbers requires only
$2N$ cells (with an alphabet of just two states) and a maximum of $2N$ updates.
Furthermore, the final state is such that a third number may be encoded onto the
ACA, and the addition process repeated. Hence the ACA may sum a whole
sequence of numbers.

There are two possible geometries for the ACA, a simple line
of cells and a closed circle. Figure 1 
shows the states through which a 10 cell ACA passes as it performs an addition; a
circular geometry is assumed so that the right-most cell of each line is adjacent to
the left-most cell.
The figure indicates how two numbers, $A$ and
$B$, must be encoded to form the ACA's initial state. The subscripts indicate a
particular bit in the binary representation of the number, $A=\sum_{i=0} A_i 2^i$,
and similarly for $B$ and the output $C=A+B$. The ACA is unusual in that it has two
different tables (update rules), the `add' and the `carry' table, which are applied
alternately.  There are also two interlaced subsets of cells, the $\alpha$ set and the
$\beta$ set; whilst one set is being updated the other remains static.
The two update tables are defined in Figure 1, and the grid shows that `carry' and
`add' are applied to first one cell subgroup and then the other. This
pattern is repeated until a total of
$2N$ individual updates have been performed (for a system with $2N$ cells). The sum
is then in the indicated cells; note that it is rotated (by $N$ cells) with
respect to the input $A$. It is because of this rotation that the circular geometry has
an advantage over the linear design; the latter would require an additional $N$ 
cells so that the answer should not `fall off the end'. Regardless of the choice of 
input numbers, the answer $A+B$ always occupies the same cells after all $2N$ steps
have been performed. This assertion can easily be proved for an ACA of reasonable size
($\sim 30$ cells or less) by exhaustive computer simulation for all possible $A$ and
$B$. Readers can verify the validity for any specific $A$ and
$B$ using the interactive version of Fig. 1
\cite{web}.

From the grid in Fig. 1 it is clear why the division into two subsets of
cells is useful. On any single update the new state of a cell depends only on
its own current state and the state of a neighboring {\em static} cell, i.e. one
which is not subject to the current update. Thus the cells that are to be
updated on a given step are independent of one-another. This means that in
the physical system, each cell may change its state at a random instant
within some given time interval (this would occur for a system driven by
photonic excitation, for example). Clearly, this spread of instants would
randomize the behavior of a simple CA in which inter-dependent cells are updated
together. 

The physical implementation of the ACA requires an
array of cells, each cell having at least two stable states and being
sensitive to the state of its neighbors. One possible cell would be the bistable
double-quantum-dot, driven through its internal states by laser pulses
\cite{Obermayer}. The beam would
encompass the whole ACA and update all cells that respond to its frequency; a single
laser could drive a large number of independent ACA devices. The potential for such
structures as a realization of a CA has been considered in some detail in
Ref. \cite{Obermayer} and will be briefly summarized here. We
use the notation x-{\bf y}-z to refer to a cell in state y whose left and right
neighbors are in states x and z respectively (for a circular CA, the sequence of
cells x-{\bf y}-z runs clockwise).

Consider the basic cell as comprising a coupled pair of quantum dots as suggested
above; the low-lying single particle states are those for which the electron
is localized on one or other of the dots. The lowest energy state within each of the
two localizations are the physical representations of `1' and `0'. One dot is
assumed to be slightly smaller than the other, so that the two localizations are
non-degenerate. The energy difference is slight and the wavefunction overlap is very
small so that the rate of spontaneous decay from one dot to the other is much slower
than the total computation time. The CA is built up from such cells
simply by producing a string of them (see e.g. the upper part of Fig. 2).
There is no tunneling allowed between the cells; they `feel' the states of their
neighbors via the Coulomb interaction. The Coulomb repulsion is greater for a pair of
cells in the same state than for a pair in opposite states; the energy difference has
a $r^{-3}$ dipole-like form, where $r$ is the cell-cell separation. If the distance to
a cell's neighbor on the right is not equal to the distance to its left neighbor, then
the two stable energy levels for an isolated cell split into eight levels, one
for each combination of neighbor states. This is shown in Figure 3. The idea of
having two species of cell and updating them alternately is realized by using two
different sizes of double-dot. The size difference shifts
the double-dot's energy levels and thus makes it possible to address one type at
a time. 

In order to produce
the desired updates of the CA, a third transient state is employed. This is a
single-particle state in which the electron probability distribution is spread
over both dots. The transient state spontaneously decays very quickly into one of
the stable states. As shown in Fig. 3, an update is produced by pumping the
system with light of a frequency that will excite cells of one size from a given
stable state (say 1-{\bf 0}-0) into the transient state. The cell may decay from
the transient state into either the original state (1-{\bf 0}-0) or the flip state
(1-{\bf 1}-0). However, if the former occurs the electron will be re-excited by
the pump, so that we can  flip the state with any desired certainty ($<1$) simply
by using a pulse of sufficient duration. Note that the frequency width of the pulse
must be sufficiently narrow to excite from only one of the levels shown in Fig. 3,
yet sufficiently broad to cover the sub-splitting (shown shaded gray) due to
non-neighboring cells.

A difficulty arises with the above system which is associated with the
bottom two lines in the `add' table of Fig. 1. The states X-{\bf 1}-0 and X-{\bf
1}-1 (where $X=0$ or $1$) transform {\em into each other}; this swap cannot be
directly translated into a sequence of pulses, as required by the physical process
described above
\cite{Ober_noted}. We must make one of the following
modifications: use $N$ extra cells of a third size, use triple rather than
double dots, employ next-nearest neighbor interactions, or use a coherent
switching process. These possibilities will be elaborated upon elsewhere
\cite{full_analysis}. Here we will focus on the first solution as it seems
the least difficult experimentally.

We will modify the system proposed in Ref. \cite{Obermayer} by using three rather
than two sizes of cell and correspondingly three inter-cell distances. A system
of this kind is shown schematically in Fig. 2. We will also need a third
update rule; this rule is
`shift'. Its effect is simply to make the target cell's state equal to the state
of its immediate clockwise neighbor. In the lower part of Fig. 2 we tabulate the
stages the ACA evolves through in terms of the light pulses to which it is
subjected. The notation
$\omega_{7\ 5}^\alpha$ denotes a pulse of the correct frequency to pump the cells
of size $\alpha$ from the state ``7"$=1-{\bf 1}-1$, via the transient state to
``5"$=1-{\bf 0}-1$. The letters C, S, and A denote `carry', `shift' and `add'
operations respectively. At first glance the grids in Figs. 1 and 2 seem quite
different, however the latter actually contains the former. Removing the shaded
squares in Fig. 2 recovers the grid in Fig. 1 (excluding the first line). Thus
$2N$ update steps in the abstract ACA correspond to $8N$ light pulses in this
chosen physical implementation. If we assume a pulse duration of $100\omega^{-1}$
\cite{Obermayer}, then a few picoseconds would be required to add two 8 bit numbers.
It is important to note that the proposed device, being composed of bistable units, is
a nanometer-scale classical computer rather than a true quantum computer. It does not
need to maintain wavefunction coherence, and is therefore far less delicate than a
quantum computer.

Possible
methods exists for reading and writing the states on the cells in parallel. One might
exploit recent experimental work
\cite{cambridge} in which fluctuations of $\pm 1$ in the number of electrons on a
quantum dot were measured using the pico-amp current flowing through a nearby
constriction. Alternatively one might employ single-electron transistors; these 
devices have recently been made to work at room temperature \cite{matsumoto}. We
expect it will prove possible to use one such technique to read all the states in
the ACA at once (by having one current probe for each ACA cell, as shown in Fig. 2).
By careful manipulation of the potential in the probe elements, it should prove
possible to use them to write data onto the ACA device as well
\cite{full_analysis}. We stress again that the the double quantum dot system
described above represents only one possible implementation; quite different systems
(for example, the Single-Electron Parametron \cite{parametron}) could prove equally
well suited.

A modest structure of just two or three cells and probe elements
would suffice to test the principles upon which the full ACA is based. In particular,
it is important to experimentally measure the various time scales involved. If
unwanted spontaneous decay should prove to be an obstacle, a solution would be to
embed the ACA device in a photonic band gap material \cite{photonic}. The ACA may
also be kept `on-course' through the addition process by making measurements of
the ACA's state {\em during} the computation. Each measurement collapses the quantum
state of the cells into `1' or `0', thus preventing a slow drift from the intended
evolution \cite{zeno}.

This work was funded by an EPSRC Photonic Materials Grant.

\newpage

\bigskip

\centerline{\bf Figure Captions}

\bigskip

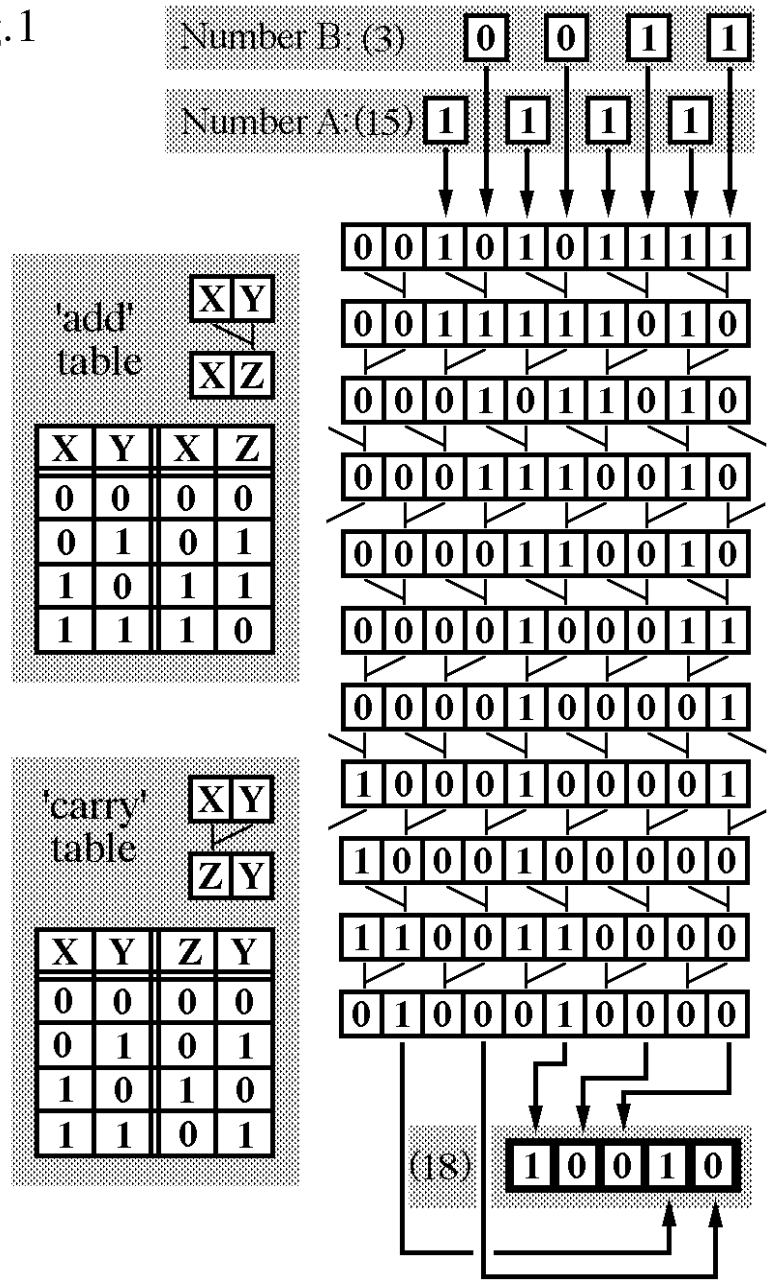
\noindent Figure 1. The `add' and `carry' tables for the ACA, together with a grid
showing the example of $15+3$. Circular
geometry is assumed, hence right-most column is adjacent to the left-most
column. See Ref. \cite{web} for an interactive version.

\bigskip
\noindent Figure 2. Above: Possible physical realization of the circular ACA
using double quantum dots. Radial lines represent channels for a probe
current, with a quantum constriction near each double-dot. 
Table: Second line shows how two 4-bit numbers, $A$ and $B$, must
be loaded into initial state of ACA. Body of table shows the
example $A=15$, $B=3$. Deleting the shaded squares recovers fundamental
ACA table in Fig. 2.

\bigskip
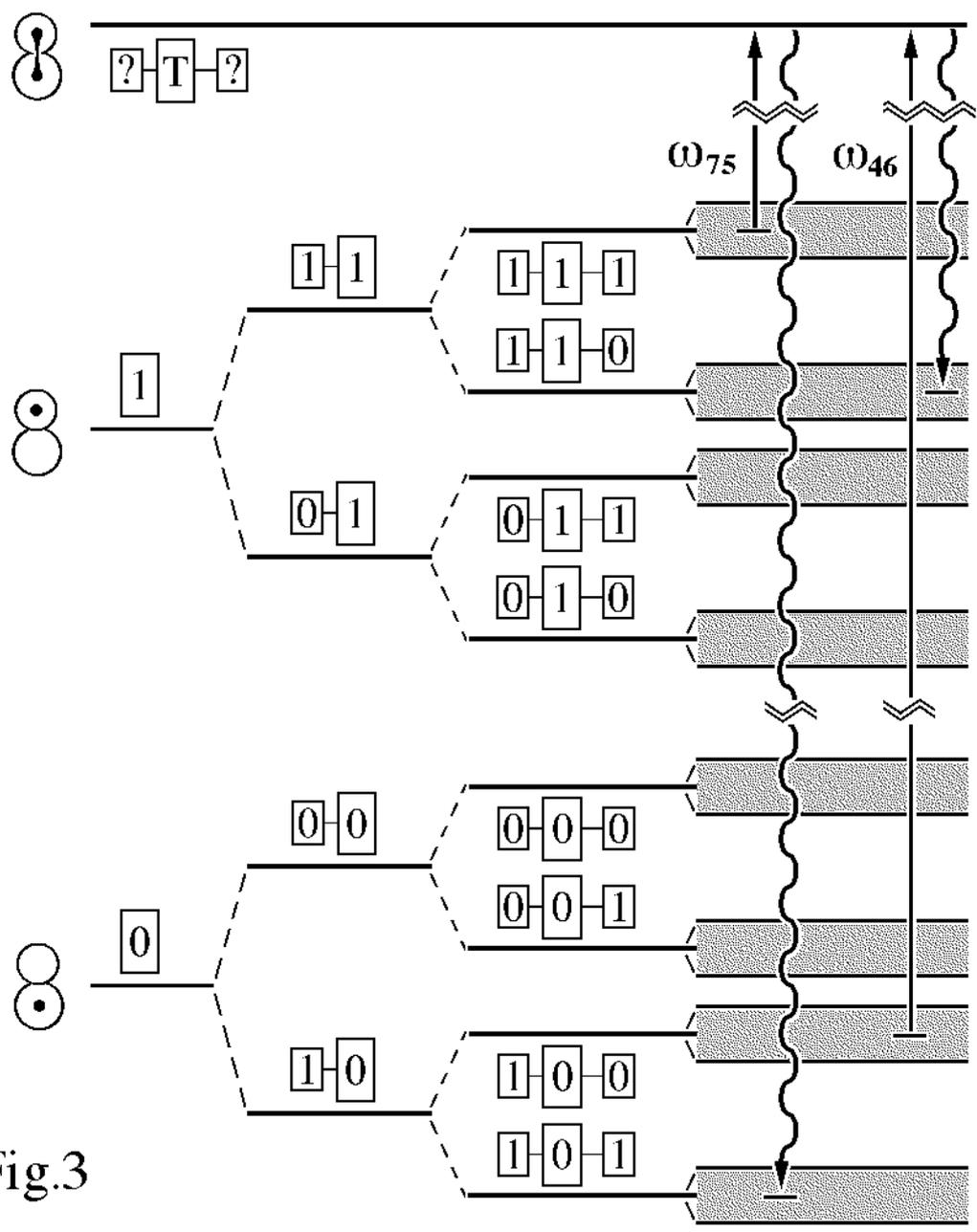
\noindent Figure 3. Left side: Energies of the `0' and
`1' states of the double quantum dot, and of the transient state `T' (electron
localization is shown schematically). Middle: Splitting of the
levels as the states of the neighboring cells are resolved.
Right side: Shaded bands show splitting due to non-adjacent cells. Vertical
lines show implementation of `add' operation; straight lines: excitation
by laser, wiggly lines: spontaneous decay.


\begin{thebibliography}{99} 

\bibitem{domino}
P. D. Tougaw and C. S. Lent, J. Appl. Phys., {\bf 75} 1818 (1994). 
\bibitem{lines}
A. N. Korotkov, Appl. Phys. Lett. {\bf 67} 2412 (1995).
\bibitem{Obermayer}
K. Obermayer, W. G. Teich and G. Mahler, Phys. Rev. B {\bf 37} 8096 and 8111
(1988).
\bibitem{parametron}
K. K. Likharev and A. N. Korotkov, Science {\bf 273} 763 (1996).
\bibitem{vonN}
J. von Neumann, {\em Theory of Self-Reproducing Automata}, Univ. of Illinois Press
(1966).
\bibitem{web}
The web page ``http://cm-th.physics.ox.ac.uk/SimonB/adder/add.html" allows
Java enabled browsers to generate Fig.1 with any desired $A$
and $B$.
\bibitem{Ober_noted}
This restriction was noted in Ref. \cite{Obermayer}.
\bibitem{full_analysis}
S. C. Benjamin and N. F. Johnson, (unpublished).
\bibitem{cambridge}
M. Field, C.G. Smith, M. Pepper, D.A. Ritchie, J.E. Frost, G.A.C. Jones and D.G.
Hasko, Phys. Rev. Lett. {\bf 70} 1311 (1993).
\bibitem{matsumoto}
K.Matsumoto, M. Ishii, K. Segawa, and Y. Oka, App. Phys. Lett. {\bf 68} 34 (1996).
\bibitem{photonic}
P.M. Hui and N.F. Johnson, {\em Solid State Physics Vol.49}, edited by H.
Ehrenreich and F. Spaepen (Academic Press, New York, 1995) p.151.
\bibitem{zeno}
B. Misra and E.C.G. Sudershan, J. Math. Phys. {\bf 18} 756 (1977).



\end{thebibliography}
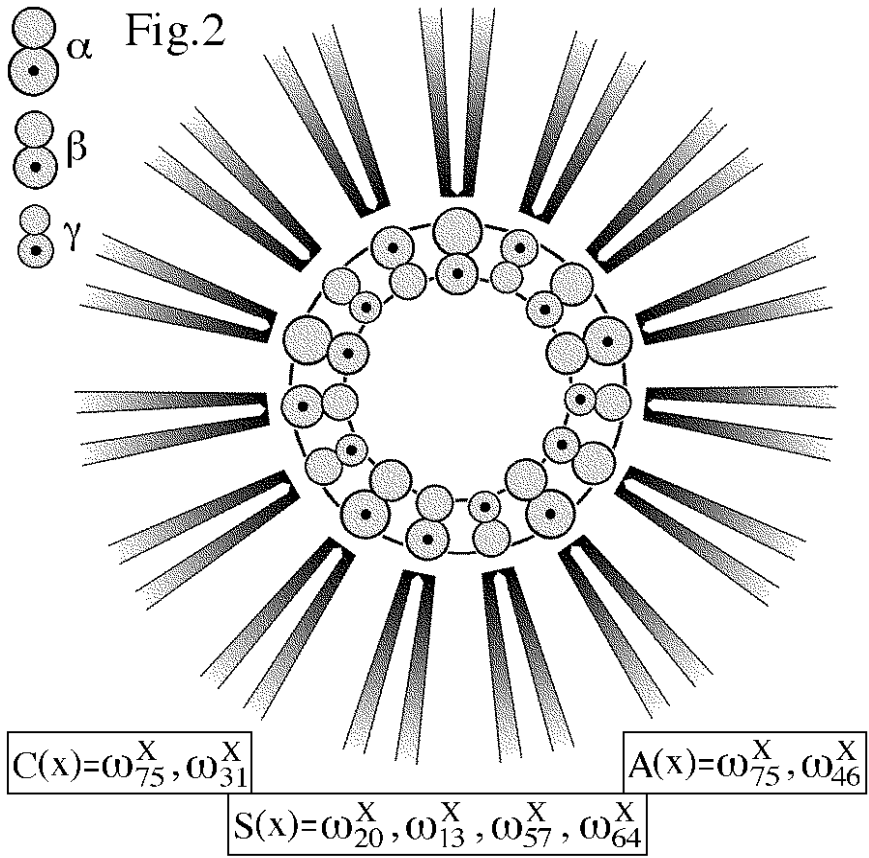
\end{document}